\documentclass[11pt]{amsart}
\usepackage{enumitem}
\usepackage{setspace}
\usepackage{geometry}                
\usepackage{natbib}
\usepackage{graphicx}
\usepackage{amssymb}
\usepackage{amsmath}
\usepackage{epstopdf}
\usepackage{esdiff}
\usepackage[linesnumbered,vlined,algoruled]{algorithm2e}
\usepackage{url}
\usepackage{mathtools}

\title{Conservation of the $t$-digest Scale Invariant}
\author{Ted Dunning}
\address{Ted Dunning \\ MapR Technologies, Inc \\ San Jose, CA}
\email{ted.dunning@gmail.com}
\date{}                                           
\begin{document}
\begin{abstract}
A $t$-digest is a compact data structure that allows estimates of quantiles which increased accuracy near $q = 0$ or $q=1$. This is done by clustering samples from $\mathbb R$ subject to a constraint that the number of points associated with any particular centroid is constrained so that the so-called $k$-size of the centroid is always $\le 1$. The $k$-size is defined using a scale function that maps quantile $q$ to index $k$. Since the centroids are real numbers, they can be ordered and thus the quantile range of a centroid can be mapped into an interval in $k$ whose size is the $k$-size of that centroid. The accuracy of quantile estimates made using a $t$-digest  depends on the invariance of this constraint even as new data is added or $t$-digests are merged. This paper provides proofs of this invariance for four practically important scale functions.
\end{abstract}
\maketitle
\section{Introduction}
The key property of the $t$-digest   \citep{t-digest-arxiv} is that centroids either have unit weight or have $k$-size less than or equal to 1. Adding data to a $t$-digest or merging $t$-digests must preserve this constraint. The $k$-size is defined using a scale function, and there are  four scale functions that are important for practical use 
\[
\begin{aligned}
k_0(q) &= \frac \delta 2 q \\
k_1(q) &= \frac \delta {2\pi}  \sin^{-1}(2q-1)   \\
k_2(q) &= \frac \delta {Z(n)} \log {\frac q {1-q}} \\
k_3(q) &= \frac \delta {Z(n)}\begin{cases}
\quad \log 2q & \text{if  } q \le 1/2 \\
- \log 2(1-q) & \text{if  } q > 1/2
\end{cases}
\end{aligned}
\]
These scale functions apply progressively more emphasis on the accuracy of quantile estimates near the tails and less emphasis on estimates near $q=1/2$. 
 
The size constraint says that any centroid with weight greater than one that spans from quantile $q_1$ to $q_2$ has a $k$-size  limited by
\[
k(q_2) - k(q_1) \le 1
\]
The addition of new data to a $t$-digest proceeds by the addition of one or more new centroids followed by the merging of consecutive centroids that have combined $k$-size subject to the size constraint. The merging cannot cause a violation of the constraint, but it is conceivable that the addition of new centroids could indirectly cause a violation due to the shifting of $q_1$ and $q_2$.

In fact, however, no such violation occurs, at least not for the scale functions above. This paper contains proofs of the preservation of this constraint as new data is added to the $t$-digest for each of these four scale functions.
\section{Proofs}
In all cases, we take $n$ as the total number of sample so far, $q_1 = n_1 / n$,  and $q_2 = n_2/n$. The centroids to the left of the one we are considering therefore have a total weight of $n_1$ and our centroid has a weight of $n_2-n_1$. For each of the scale functions, we need to consider the addition of $\Delta n$ samples either to the right or left of our centroid. For some functions, addition to the right and left are logically the same due to symmetry and thus only one alternative need be considered.
\subsection{For $k_0$}
The function $k_0$ is completely symmetric  so we need only consider the case of data added to the right of the centroid in question. Subsequent to this addition, the $k$-size of our centroid will become
\begin{align*}
k_0(q'_2)-k_0(q'_1) &=\frac \delta {2(n+\Delta n)}\left ( n_2 - n_1 \right) \\
&< \frac \delta {2n} \left ( n_2 - n_1 \right)=k_0(q_2) - k_0(q_1) \le 1 
\end{align*}
This means that the $k$-size of a centroid can only shrink with the addition of new data and thus, the constraint on maximum $k$-size is preserved for $k_0$.
$\square$
\subsection{For $k_1$}
The $k_1$ scaling function is symmetric so we only have to consider the case of new data to the right of the centroid of interest. 

Note that
\begin{align*}
\diff {\left[k_1(q_2)-k_1(q_1)\right]} {n} &= \diff {k_1(q_2)}{n} -\diff { k_1(q_1)} {n} \\
&= -\frac {1} n \left( \sqrt{\frac {n_2} {n-{n_2}}} - \sqrt{\frac {n_1} {n-{n_1}}} \,\right)
\end{align*}
This is uniformly negative if $n > n_2 > n_1\ge 0$ which implies that any increase in $n$ will result in a decrease in the $k$-size of a centroid that spans $q_1 = n_1/n$ to $q_2 = n_2/n$. The constraint is thus preserved for $k_1$.
$ \square$
\subsection{For $k_2$}
For the scaling function $k_2$ the same symmetry consideration applies as for $k_1$ so we only have to consider the addition of data to the right of the centroid of interest. Moreover, $k_2(q)$ increases without bound as $q \rightarrow 0$ and $q \rightarrow 1$ so the first and the last centroids must have unit weight and  need not be considered (so $0 < q_1<q_2<1$). The new $k$-size of some remaining centroid will be
\begin{align*}
k_2(q'_2)-k_2(q'_1)&= \frac \delta {Z(n+\Delta n)} \left( \log \frac {n_2} {n-n_2+\Delta n}-\log \frac {n_1} {n-n_1+\Delta n} \right) \\
&= \frac \delta {Z(n+\Delta n)} \left( \log \frac {n_2} {n_1} + \log \frac  {n-n_1+\Delta n}{n-n_2+\Delta n} \right) \\
\end{align*}
But $Z(n)$ is non-decreasing and we know that $n>n_2 > n_1 > 0$, so $n-n_2 < n-n_1$ and
\begin{align*}
\frac  {n-n_1+\Delta n}{n-n_2+\Delta n} <  \frac  {n-n_1}{n-n_2}
\end{align*}
So,
\begin{align*}
k_2(q'_2)-k_2(q'_1) &< \frac \delta {Z(n)} \left( \log \frac {n_2} {n_1} + \log \frac  {n-n_1}{n-n_2} \right) =k_2(q_2)-k_2(q_1) \le 1
\end{align*}
Thus
\begin{align*}
k_2(q'_2)-k_2(q'_1) &< k_2(q_2)-k_2(q_1) \le 1
\end{align*}
and the constraint is preserved for $k_2$.
$\square$
\subsection{For $k_3$}
The definition of $k_3$ has two branches
\begin{align*}
k_3(q) &= \frac \delta {Z(n)}\begin{cases}
\quad \log 2q & \text{if  } q \le 1/2 \\
- \log 2(1-q) & \text{if  } q > 1/2
\end{cases}
\end{align*}
Except where $q_1<1/2<q_2$, we know by symmetry that we only need to consider the first of these branches, but there are two cases within that branch where the new samples are either to the right of the centroid of interest or to the left.

For the case where data is added to the right, $n_1$ and $n_2$ remain unchanged and since the denominator in $q'_i = n_i/(n+\Delta n)$ cancels out, the $k$-size is unchanged except for the normalizing factor $Z(n)$ which is non-decreasing.
\begin{align*}
k_3(q'_2)-k_3(q'_1) &= \frac {\delta} {Z(n+\Delta)} \left(\log  \frac {n_2} {n+\Delta n} - \log \frac {n_1}{n+\Delta n}  \right) \\
 &= \frac {\delta} {Z(n+\Delta)} \left(\log  \frac {n_2} {n_1} \right) \\
&\le\frac {\delta} {Z(n)} \left(\log  \frac {n_2} {n_1} \right)= k_3(q_2)-k_3(q_1)\le 1
\end{align*}

For the case where data is added to the left, start with the fact $0 < n_1 < n_2$. We know
\begin{align*}
\frac {n_2 + \Delta n} {n_1 + \Delta n} &< \frac {n_2} {n_1}
\end{align*}
From this, 
\begin{align*}
k_3(q'_2)-k_3(q'_1) &= \frac {\delta} {Z(n+\Delta n)} \left(\log  \frac {n_2+\Delta n} {n_1+\Delta n} \right) \\
&< \frac {\delta} {Z(n+\Delta n)} \left(\log  \frac {n_2} {n_1} \right) \\
&< \frac {\delta} {Z(n)} \left(\log  \frac {n_2} {n_1} \right) = k_3(q_2)-k_3(q_1) \le 1
\end{align*}
This leaves only the case where $q_1<1/2<q_2$. We need only consider the case where the new data is added to the right. Note that
\begin{align*}
k_3(q_2)-k_3(q_1) &= \frac {\delta} {Z(n)} \left(-\log \left( 1- \frac {n_2} {n} \right) - \log \frac {n_1} {n} \right) \\
 &= \frac {\delta} {Z(n)} \log \left( \frac {n^2} {n_1(n-n_2)}    \right) 
\end{align*}
For the log term, the derivative with respect to $n$ is
\[
\diff {} {n} \log \left( \frac {n^2} {n_1(n-n_2)}    \right) =\frac {n(n - 2n_2) }{n_1  (n - n_2)^2}
\]
This is uniformly negative if $n_2 > n/2$. For $n > 2 n_2$, however, we revert to the first branch of the $k_3$ and  the $k$-size no longer changes with increasing $n$ except for the effect of $Z(n)$ which we know cannot make $k$-size any larger. 

This means that for $q_1<1/2<q_2$ we know $k_3(q'_2) - k_3(q'_1) < k_3(q_2) - k_3(q_1)$.

The $t$-digest size constraint is thus preserved for $k_3$ in all cases and sub-cases.
$\square$

\bibliographystyle{chicago}
\bibliography{refs}{}

\end{document}